\documentstyle[aps,twocolumn]{revtex}
\author{\bf Mario Nicodemi and Antonio Coniglio}
\address{Dipartimento di Scienze Fisiche, Universit\`a di Napoli
``Federico II", Sezione INFM e INFN di Napoli
\\ Pad. 19 -- Mostra d'Oltremare, 80125 -- Napoli -- ITALY}
\title{Macroscopic glassy relaxations and microscopic motions in a
  Frustrated Lattice Gas}
\date{\today}
\newcommand{\be}{\begin{equation}}
\newcommand{\ee}{\end{equation}}
\newcommand{\beqs}{\begin{eqnarray*}}

\newcommand{\eeqs}{\end{eqnarray*}}
\newcommand{\bi}{\begin{itemize}}
\newcommand{\ei}{\end{itemize}}

\newcommand{\lan}{\langle}
\newcommand{\ran}{\rangle}

\begin{document}
\maketitle
\begin{abstract}
We study microscopic and macroscopic dynamical properties of 
a frustrated lattice gas showing the violation of Stokes-Einstein
law. The glassy behaviors are analyzed and
related with experimental results in glass former systems. 
\end{abstract}
\bigskip

In simple liquids the connections between the times of macroscopic
relaxations and the properties of microscopic particles motion are
elucidated by the Stokes-Einstein relation which links shear viscosity
and particle diffusivity. Recent experimental evidence shows how such
a relation is violated in supercooled glass forming liquids.
In these systems complex dynamical behaviors, which are 
distributed on many time scales, 
are supposed to be linked in a non trivial way to atomic motions 
of the liquid components \cite{GS-Rev,Angell-Science95}. 

We face these questions in the context of a microscopic model recently
introduced to describe some general features of glasses \cite{NC}.
This model, which bridges Spin Glasses (SG) 
\cite{BinderYoungRev} and site 
Frustrated Percolation \cite{ConiglioFP}, consists in a 
lattice gas model in which each particle contains an internal degree of freedom
characterized by a spin variables. The spins interact via quenched 
ferromagnetic and antiferromagnetic interactions randomly distributed like
in the SG model.
The model is described by the following dilute SG Hamiltonian:
\be
\beta {\cal H} = -J\sum_{\lan ij \ran} 
(\epsilon_{ij}S_iS_j - 1)n_in_j -\mu \sum_i n_i 
\label{H}
\ee
where $n_i=0,1$ are occupancy variables which have an internal Ising 
degree of 
freedom $S_i\pm 1$, the $\epsilon_{ij}=\pm 1$ are quenched random lattice
interactions and $\mu$ is an adimensional chemical potential,
which plays the role of the inverse of the temperature  1/T.

When all the sites are occupied ($n_i=1$), i.e. if $\mu\rightarrow\infty$,
this model is the standard $\pm J$ Ising Spin Glass.  
In the limit, $J\rightarrow\infty$, 
it describes a frustrated lattice gas where two
nearest neighbor particles can be occupied only if their internal variables 
satisfy the constraint $\epsilon_{ij}S_iS_j = 1$. 
Since in a frustrated interaction loop
the spins cannot satisfy all the couplings, in this model particle 
configurations in which a frustrated loop \cite{BinderYoungRev} 
is fully occupied are not allowed .

In a liquid the internal degrees of freedom $S_i$ may be associated 
for example to the internal orientation of particles, but, 
more generally, they represent microscopic quantities which probe
local effects of ``frustration". 
Geometrical hindrance in the liquid implies the existence of loops which 
cannot be fully occupied, in correspondence with the frustrated loops in 
the model (see \cite{Rivier}).

In mean field\cite{Arenzon} the static properties of the model
show two transitions: one at a value $\mu_c$ 
which signals the appearance of 
metastable states and a second thermodynamic transition at
a higher value $\mu_0$ (corresponding to a lower temperature).
These properties recall those of the $p$-Spin Glass models\cite{CuKu}.
These models have received renewed attention, as their dynamical behaviour
in mean field is described by the same equation as those of the mode
coupling theory for simple liquids\cite{GS-Rev}. Moreover the dynamical 
transition coincide with the first  transition from the stable to
the metastable state.
The similarity in the static properties of the two models suggest that 
also the dynamics may be the same. Therefore we 
expect for the lattice gas model ~(\ref{H}) in mean field a
dynamical transition at $\mu_c$  with a
diffusion coefficient vanishing 
with a power law.

We have studied the model~(\ref{H}) in $d = 3$ dimensions in the limit 
$J\rightarrow\infty$ by means of Monte Carlo methods. 
We use a standard Monte Carlo Dynamics, in which particles diffuse 
and spins are updated with Metropolis spin flip.
Some results
on the static and dynamic properties of the model have been previously reported
\cite{NC}. Here we focus mainly on the relation 
between the diffusion coefficient and the characteristic relaxation time.

Our data are consistent with the following picture:
we find a region  $\mu < \mu_p$ (high temperature)
where the model behaves as a normal fluid characterized by single exponential
relaxation and  normal diffusion. The value of $\mu_p$ is numerically 
consistent with
the percolation threshold for the particles system. For values 
$\mu_p < \mu < \mu_0$ the density density time dependent
autocorrelation function exhibits a two step relaxation, corresponding
to the $\alpha$ and $\beta$ relaxation times. The value 
$\mu_0$ corresponds to the spin glass transition characterized
by the divergence of the non linear susceptibility
and where moreover the diffusion coefficient, $D(\mu)$, vanishes
\cite{NC}. In this region $D(\mu)$ exhibits a net cusp at $\mu^*$
($\mu_p < \mu^* < \mu_0$). 
Both the diffusion coefficient and the $\alpha$ relaxation time
exhibit a power law behavior for $\mu < \mu^*$ 
with a sharp crossover to 
an Arrhenius or 
Vogel-Fulcher law for $\mu > \mu^*$.
This crossover can be interpreted in the following way \cite{parisi}. 
In mean field 
the dynamical transition occurs at the same temperature
where the static exhibits a transition from a stable to a metastable state.
Once in the metastable state the system is trapped forever due to the
infinite lifetime of
the mean field metastable state. In finite dimension the metastable state 
has a finite lifetime,
and the system does not reach a full arrest.

The value of $\mu_p$ is located roughly where
the $\beta$ relaxation and the $\alpha$ relaxation time starts to separate and
where the relation between the 
$\alpha$ relaxation time and the diffusion coefficient
violates the {\em Stokes-Einstein law}, as 
observed in real structural glasses 
\cite{Leporini,Rossler,TK,Sillescu,Pusey}.

We have considered a cubic lattice of linear sizes 
$L=8,12,16$ with periodical boundary conditions,
fixing $J=10$ and using $\mu$ as a variable. 
Our results do not change for $J=10000$, showing that we are 
effectively close to the value of $J = \infty$.
As stated, the {\em dynamic} in our model 
is a standard Monte Carlo (MC) dynamics, in which particle diffuse and spins  
are updated with Metropolis spin flip. The system is equilibrated 
after successive thermalization at higher and higher values of $\mu$ for 
about $10^6$ MC steps at each fixed value of external parameters. 
The measures are then taken up to about $10^{8}$ MC steps, for a given random 
configuration of the couplings $\epsilon_{ij}$. 
We are aware that in the deep glassy region (above $\mu \sim 5$) our 
numerical results may be only indicative due to the required very 
long simulation times, but this do not change the overall picture 
described above. 

We have calculated the density-density  time dependent autocorrelation 
functions:
\begin{equation}
C_k(t)= \lan \rho_k(t)\rho_{-k}(0) \ran / \lan \rho_k\rho_{-k}\ran
\label{corden}
\end{equation}
Here $\rho_k$ is the Fourier transform on the lattice of the density 
$\rho_k(t)= \frac{1}{N} \sum_{r=0}^{L-1} \rho(r,t) \cos(q_k r)$, where 
$\rho(r,t)$ is the particle density, at time $t$ and at a distance $r$ 
from a median plane in our cubic lattice of size $L$, 
and $q_k=\frac{2\pi}{L}k$, with $k\in\{1,2,...,L/2\}$. 
Times are measured in such a way that $t=1$ corresponds on average to 
the time to update once all the degree of freedom in the system. 
We have also calculate the the square magnetization time dependent 
autocorrelation function and found that behaves in a similar way.

\begin{figure}[h]
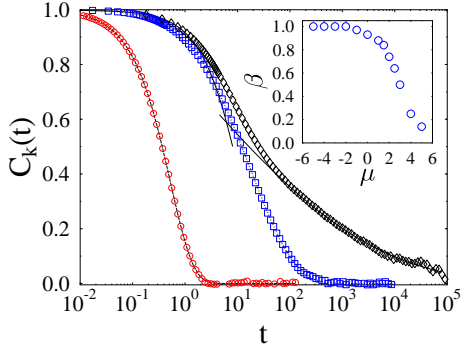

\caption{ 
Fourier transformed density correlation function, 
$C_k(t)$, for $k=L/4$, as a function of time $t$ for several values of 
the chemical potential $\mu$ ($\mu=-2.0,3.0,5.0$), in a system of linear 
size $L=16$ and coupling constant $J=10$. 
Superimposed are the short time exponential and the long time 
stretched exponential fits (see text).
Inset: 
The exponent $\beta$ of the stretched exponentials of the long time fits of 
density correlation functions reported in the main frame, 
as a function of $\mu$.
}
\label{corr_k}
\end{figure}

As anticipated above, the system at low $\mu$ (i.e. low density) behaves as 
normal liquid, 
with exponentially decaying time correlation functions
(see Fig.~\ref{corr_k}): $C_k(t)\sim \exp(-t/\tau_0(k))$.
The characteristic time $\tau_0(k)$, excluding finite size effects,
is inversely proportional to $k^2$, 
$\tau_0(k) \sim 1 / D_0 k^2 + \tau_{\infty}$. 

For larger values of $\mu$ (larger density) even if the short time decay ($t\le 1$) is still 
exponential, the long time relaxation ($t>>1$) 
may be reasonably fitted by the Kohlrausch-Williams-Watt 
stretched exponential form (see Fig.~\ref{corr_k}): 
\begin{equation}
C_k(t)\sim B\exp(-(t/\tau (k))^{\beta})
\label{expstr}
\end{equation}
with $\beta$ function of $\mu$ (see Fig.~\ref{corr_k}), 
as typically occurs in SG \cite{Ogielski}. 
This implies that, the MCT prediction of 
{\em time-temperature superposition relation} \cite{GS-Rev} 
is verified only in very narrow density  
intervals. 

\begin{figure}[h]
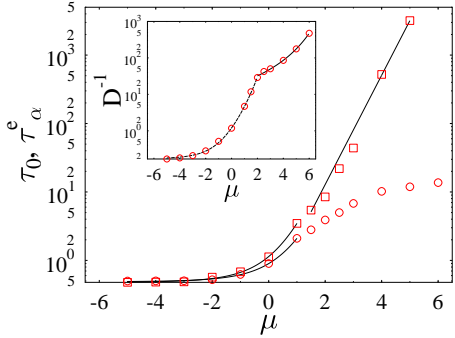

\caption{ The short time exponential $\beta$-relaxation time
  $\tau_0(k)$ (circle), 
and the $\alpha$-relaxation 
$\tau^e_{\alpha}\equiv\tau_{\alpha}/\ln(10)$ (square), of density 
correlation function
(for $k=L/4$), as a function of $\mu$, 
for a system of linear size $L=16$ and $J=10$. 
Continuous curves are power law and Arrhenius fit described 
in the text. The point where the low $\mu$ behaviors fails is located
around $\mu_p\sim 0.75$. 
Inset: 
The inverse of diffusivity, $D(\mu)$, as a function of $\mu$ 
for the same system. 
Superimposed are the power law and Voghel-Tamman-Fulcher fits described in 
the text. The cusp in $D(\mu)$ individuates a characteristic value
$\mu^*\sim 2.0$ ($\mu^*<\mu_0$, where $\mu_0$ is the SG transition).
}
\label{t0_and_t2_k3}
\end{figure}

Close to $\mu_p\sim 0.75$ 
($\rho_p\sim 0.38$), 
we observe the separation of short times
($\beta$-process) 
and long times ($\alpha$-process) relaxation. 
For an estimate of the long time relaxation,
the $\alpha$-relaxation, 
in principle we could use 
$\tau (k)$ from the fit of eq. (3). However,
due to the sensitivity of 
$\tau (k)$ on the details of the fit, such as for example 
the region chosen for the fit, following
Ref.~\cite{kob}
we define the $\alpha$ relaxation 
$\tau_{\alpha}(k;\mu)$
as the time such that $C_k(\tau_{\alpha})=10^{-1}$. 
For $\mu < \mu^* \sim 2$, $\tau_{\alpha}(k)$  
is well fitted with a power law behavior 
(see Fig.~\ref{t0_and_t2_k3}) \cite{nota_pl}:
\begin{equation}
\tau_{\alpha}(k;\mu)=A_k(\mu_c-\mu)^{-\gamma}+B_k
\label{potenza}
\end{equation}
with $\gamma=6.8$ and $\mu_c=5.6$ almost independent of $k$, 
($A_k=7000$ and $B_k=0.5$ for $k=L/4$). Also the time $\tau_0(k)$, of the 
initial exponential like decay,  
follows the same power law dependence up to $\mu_p$, and then
saturates to a finite value exponentially with $\mu$.

For $\mu > \mu^*$, while $\tau_0(k;\mu)$ 
saturates, $\tau_{\alpha}(k;\mu)$ is reasonable fitted by
an Arrhenius behavior divergence
(see Fig.~\ref{t0_and_t2_k3}):
\begin{equation} 
\tau_{\alpha}(k;\mu)=a_k ~ \exp(\mu/\mu_{\alpha})
\label{divesp}
\end{equation}
with $a_k=0.7$ and $\mu_{\alpha}=0.4$ for $k=L/4$.

The crossover from a power law behavior forecasted by MCT or Spin
Glass mean field theory to an 
Arrhenius (or Vogel-Tamman-Fulcher in fragile liquids) behavior 
for $\alpha$ relaxation times, is typically observed in real structural  
glasses too \cite{Angell-Science95}.

\begin{figure}[h]
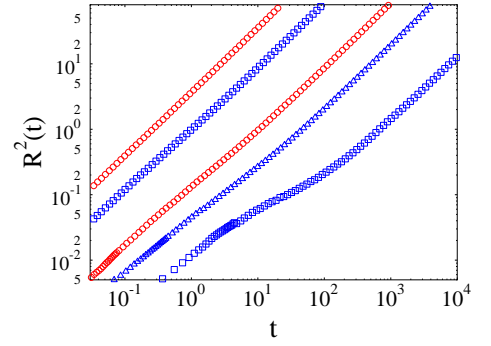

\caption{Particles mean square displacement, $R(t)^2$, as a function of time 
in a system of linear size $L=16$, for $J=10$, and for 
$\rho= 0.088, 0.271, 0.440, 0.581, 0.674$
(higher curves correspond to lower densities).}
\label{msd_16}
\end{figure}

It is interesting to compare the anomalies of the autocorrelation functions 
described above, to microscopic particle diffusion.
Here frustration may have a strong effect on particles.
We have calculated the particle mean square displacement \cite{NC}, 
$ R^2(t)= \lan \frac{1}{N}\sum_i (r_i(t)-r_i(0))^2 \ran $.
For $\mu < \mu_p $ we find a Brownian typical linear time behavior. For 
$\mu > \mu_p$ , in the intermediate time region, 
we observe a subdiffusive region with an inflection 
which becomes more evident as $\mu $ increases 
(see Fig.\ref{msd_16}).

The linear asymptotic behavior of $R^2(t)$, defines the diffusivity $D$. 
This shows an apparent cusp at $\mu^*\sim 2$ ($\rho^*\sim 0.5$) 
and an abrupt change in behavior, as shown in Fig.\ref{t0_and_t2_k3}. 
Below $\mu^*$, it is possible to fit $D(\mu)$ with the power law given
in eq.(\ref{potenza}) with the same $\gamma$ and $\mu_c$ 
(see Fig.~\ref{t0_and_t2_k3}) found studying density relaxation times. 

The value $\mu_c$ from the power law fit 
corresponds to the characteristic temperature $T_c$ of 
Mode Coupling Theory \cite{GS-Rev}, or to the ``dynamic transition" of the 
mean-field theory of $p$-Spin Glasses \cite{BinderYoungRev,CuKu}. 
Above $\mu^*$ the best fit for $D$ is obtained using a Voghel-Tamman-Fulcher 
(see Fig.~\ref{t0_and_t2_k3}):

\begin{equation} 
D^{-1}(\mu)=A_D ~ \exp(d/(\mu^{-1}-\mu_1^{-1}))
 \label{VFT}
\end{equation}
with $A_D=17$, $\mu_1=11.4$ and $d=0.3$, for the system of size $16^3$. 
However an Arrhenius fit  (i.e. 
$\mu_1\rightarrow\infty$), as that found for 
$\tau_{\alpha}$ of $\alpha$ relaxation, is just slightly worse:
\begin{equation} 
D^{-1}(\mu)=A_D' \exp(\mu/\mu_{D})
\label{Arrhenius}
\end{equation}
where $A_D'=2.72$ and $\mu_{D}=1.1$.  
A crossover from power law to Arrhenius (or Voghel-Tamman-Fulcher) 
behavior is also observed in real experiments 
\cite{Angell-Science95}. 

In our model the density $\rho^*$ corresponding to 
$\mu^*$, where the Arrhenius  region sets in, is 
signaled by a local minimum in the static structure factor 
$S(k;\rho)$, as a function of $\rho$.
Our observations indicate that the diffusivity $D$ goes to zero in the region
where the static SG transition, $\mu_0$, should be located as signaled
by the divergent non linear susceptibility $\chi_{SG}$. 
However it is difficult to numerically
establish where exactly $\mu_0$ is located ($\mu_0 \geq 5.5$) \cite{NC}. 
As in real
glasses, however there is no divergence of density fluctuations.

\begin{figure}[h]
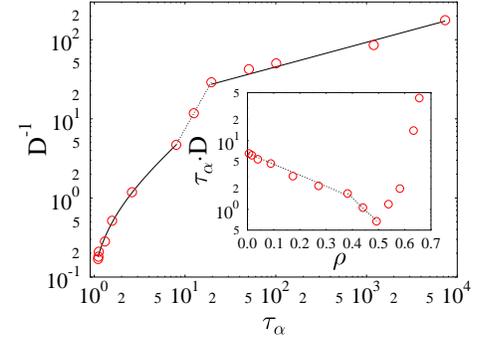

\caption{The diffusivity, $D$, as a function of the 
$\alpha$-relaxation time, $\tau_{\alpha}$. 
As in real experiments three different regions appear.
The superimposed curves are fit with the generalized fractional
Stokes-Einstein law $D^{-1}\sim \tau_{\alpha}^{\xi}$. 
Inset:
The product of diffusivity, $D(\rho)$, and of $\alpha$-relaxation
time, $\tau_{\alpha}(\rho)$, as a function of density $\rho$. 
}
\label{d_vs_t2_k3}
\end{figure}

We compare now the diffusion coefficient D and the 
$\alpha$ relaxation time $\tau_{\alpha}$. 
In a normal liquid the diffusion coefficient and the  viscosity, $\eta$, 
are related by the 
Stokes-Einstein relation (SE):
$\eta D = C T$ 
where $T$ is the temperature in Kelvin and $C$ a constant \cite{nota2}. 
In our system it is not possible to directly define a viscosity, 
but due to the Maxwell relation $\eta$ is proportional to the time scale of 
the asymptotic 
relaxation $\tau_{\alpha}$. We find that the product 
$\tau_{\alpha}  D$ is not a 
constant varying the potential $\mu$,  as shown in 
in Fig.~\ref{d_vs_t2_k3}. Studying $\tau_{\alpha}(\mu)  D(\mu)$, three 
regions emerge, separated by the values of 
$\mu_p$ and $\mu^*$. 
A similar separation in three 
different regions is found in real experiments on colloidal suspensions
\cite{Pusey}, where, 
plotting the product $\eta D$ as a function of the volume 
fraction, a behaviour is found close to that depicted in the inset 
of our Fig.~\ref{d_vs_t2_k3}.

It has been suggested that the departure from the Stokes Einstein relation, 
in glass forming liquids\cite{Leporini},  could be described by the following
relation:
$D^{-1} = K\eta^\xi$.
To check if such relation holds in our model we have plot 
(see Fig.~\ref{d_vs_t2_k3})
 $1/D$ as a 
function of $\tau_{\alpha}(k=L/4)$. By  assuming
\begin{equation}
D^{-1} = K\tau_{\alpha}^\xi
\label{DSEgen}
\end{equation}
we find  for $\mu\le \mu_p$ 
$\xi=1$ (as in usual SE), and for 
$\mu\ge\mu^*$ $\xi\sim 0.3$. 
This last value of $\xi$ for $\mu\ge\mu^*$ is consistent 
with eq.~(\ref{divesp}) 
and eq.~(\ref{Arrhenius}) and $\xi=\mu_{\alpha}/\mu_D$ 

Experiments in glass forming liquids \cite{Leporini,Rossler,TK,Sillescu}
indicates that the exponent is $\xi\leq 1$, ranging in a broad 
interval depending on the system (for instance in
$o$-terphenyl using the rotational diffusion coefficient 
is found $\xi=0.28$ \cite{Leporini}, in PMMA $\xi=0.69$ 
\cite{Sillescu}). 
It is striking that the data on $o$-terphenyl\cite{Leporini}, not only exhibit
the same  exponent $\xi$ found here, 
but the rotational diffusion coefficient
also shows a cusp as function of $1/T$ and 
the appearance of several regions, 
analogous to those depicted respectively in Fig.2 and Fig.~\ref{d_vs_t2_k3}. 

In conclusion 
we have studied the macroscopic relaxation and microscopic diffusive 
properties 
of the frustrated Ising lattice gas introduced in \cite{NC}. Many connections have appeared with the physics of real glass forming
liquids ranging from the anomalies in the density relaxations, to 
those of diffusivity, to violation of Stokes-Einstein law 
\cite{GS-Rev,Angell-Science95}.

The authors are grateful to Dino Leporini for useful discussions.
This research has been partially supported by a grant of CNR.

\begin{thebibliography}{50}
\bibitem{nota_pl} Actually other fit functions may be used. We reported 
the power law fit because it is slightly better than other fit such as 
exponential or Vogel-Fulcher 
\cite{NC_futuro}. 
Moreover the power law fit is consistent with the mean field picture
of a power law behaviour at a dynamical transition above the static transition.
\bibitem{nota2} Note that the Stokes-Einstein law has 
a general validity only in the hydrodynamical 
limit, namely the 
size of the diffusive particle must be very large
compared with the molecular size of the liquid. 
Experimentally in normal liquids the SE  law is verified 
even for particle sizes of the order of the molecular size. For this
range of particle sizes  the SE law must be considered a phenomenological law,
which seems to break down near the glass transition.
This is also the case for our model where the diffusive particle
coincides with the particle fluid.
\end{thebibliography}
\end{document}